\documentclass[12pt,preprint]{aastex}
\usepackage{emulateapj5}
\usepackage{graphicx}
\usepackage{epsfig}

\newcommand{\RS}{{\mathrm R}_{\mathrm{S}}}

\lefthead{MANCHESTER ET AL.}
\righthead{Simulation of the October 28, 2003 CME}

\received{\today}
\revised{???}
\accepted{???}

\authoraddr{
Center for Space Environment Modeling, University of Michigan,
Ann Arbor, Michigan, 48109-2143.
(chipm@umich.edu)}

\begin{document}

\title{Three-Dimensional MHD Simulation of the 2003 October 28 Coronal Mass Ejection: Comparison with LASCO Coronagraph Observations}

 \author{
  Ward B.~Manchester IV,\altaffilmark{1} Angelos Vourlidas,\altaffilmark{2} 
  G{\'a}bor T{\'o}th,\altaffilmark{1}    No{\'e} Lugaz,\altaffilmark{3}
  Ilia I.~Roussev,\altaffilmark{3}       Igor V.~Sokolov,\altaffilmark{1}
  Tamas I.~Gombosi\altaffilmark{1}       Darren L. De Zeeuw\altaffilmark{1}
  Merav Opher\altaffilmark{4}}


\altaffiltext{1}{Center for Space Environment Modeling, University of Michigan, Ann Arbor, Michigan}
\altaffiltext{2}{Naval Researach Laboratory, Washington D.C.}
\altaffiltext{3}{Institute For Astronomy, University of Hawaii, Honolulu, Hawaii}
\altaffiltext{4}{George Mason University, Fairfax, Virginia}

%
%

\begin{abstract}
We numerically model the coronal mass ejection (CME) event of 
October 28, 2003 that erupted
from active region 10486 and propagated to Earth in less than
20 hours causing severe geomagnetic storms.  The magnetohydrodynamic 
(MHD) model is formulated by first arriving at a steady 
state corona and solar wind employing synoptic magnetograms.
We initiate two CMEs from the same active region, one approximately 
a day earlier that preconditions the solar wind for the much faster
CME on the 28th.  This second CME travels through the corona at
a rate of over 2500 km s$^{-1}$ driving a strong forward shock.  We 
clearly identify this shock in an image produced by the Large Angle 
Spectrometric Coronagraph (LASCO) C3, and reproduce the shock and
its appearance in synthetic white light images from the simulation.
We find excellent agreement with both the general morphology and  
the quantitative brightness of the model CME with LASCO observations.  
These results demonstrate that the CME shape is largely determined
by its interaction with the ambient solar wind and may not
be sensitive to the initiation process.  We then show how the
CME would appear as observed by wide-angle coronagraphs
onboard the Solar Terrestrial Relations Observatory (STEREO) 
spacecraft.  We find complex time evolution of the
white-light images as a result of the way in which the
density structures pass through the Thomson sphere.  The simulation 
is performed with the Space Weather Modeling Framework (SWMF). 
\end{abstract}


\section{Introduction}
The October-November period of 2003 saw some of the most energetic solar  
flares and coronal mass ejections of any solar cycle.  These eruptive events
came to be known as the Halloween Events and included 11 X-class flares,
6 radiation storms and 4 geomagnetic storms.  These eruptions originated
from three active regions (ARs), 10484, 10486 and 10488 of which 10486 was
the largest, most complex and most active producing 12 major coronal
mass ejections (CMEs).  Three of these CMEs that occurred on October 28,
29 and November 4 were particularly energetic with speeds in excess of 
2000 km s$^{-1}$ and were associated with X17, X10 and X28 flares respectively. 
The October 28 CME was very geoeffective and damaged satellites, diverted 
airplane routes, caused power failures in Sweden, disrupted 
long-distance radio communications, and caused northern lights 
(aurora borealis) as far south as Florida.  The November 4 CME
was the most energetic of the Halloween Events, but occurred while
AR 10486 was on the western limb so that the CME was less geoeffective
as the CME was not directed toward Earth. 

Because of their extreme nature, the Halloween Events have been the 
subject of a great deal of study, both observational as well as
theoretical and modeling efforts.  The October 28 CME in particular
was well observed from Sun to Earth beginning with the vector magnetic 
field of AR 10486 \citep[]{Liu:2005}, and surface flow measurements
\citep[]{Yang:2004, Deng:2006}.  For this event, flare loops were 
observed by TRACE \citep{Su:2006}, and the global coronal field
has been reconstructed from data taken from the Michelson Doppler
Imager (MDI) aboard the Solar and Heliospheric Observations (SOHO) 
spacecraft \citep[]{Liu:2006}.  Dense plasma associated with the October 28
CME has been observed to propagate past the Earth by the 
Solar Mass Ejection Imager (SMEI) \citep[]{Jackson:2006} and was also  
detected by interplanetary scintillation (IPS) \citep[]{Tokumaru:2007}. 
Finally, the event was observed {\it in situ} by Wind and the
Advanced Composition Explorer (ACE) \citep[]{Skoug:2004, Zurbuchen:2004},
which reveal magnetohydrodynamic (MHD) quantities in the solar wind as 
well as ion abundances.
 
Efforts to simulate the Halloween Events include those of
\citet{Intriligator:2005} who modeled the propagation of the CME
plasma from 2.5 $\RS$ to 10 AU with a three-dimensional (3D) kinematic model. 
\citet{Krall:2006} modeled the 2003 October 28-30 period with a 
one-and-a-half-dimensional (1.5D) reduced MHD model that describes
the self-similar expansion of a magnetic flux rope.  In this case,
the authors also coupled the heliospheric results 
with a 3D magnetosphere code and compared it with simulations
driven by the observed solar wind data. \citet{Liu:2006} modeled
the CME eruption in the solar corona with density and pressure
perturbations.  \citet{Dryer:2004} and \citet{Wu:2005} both modeled 
CME-driven shock propagation and predicted the shock arrival times
with reasonable success.  This work to model a specific event marks a 
clear distinction from non-event-specific CME simulations  
\citep[e.g.][]{Usmanov:1995, Riley:2002, Manchester:2004b, Jacobs:2007},
and non-event-specific Sun-to-thermosphere space weather simulations 
modeled by the Center for Space Environment Modeling (CSEM) 
\citep[]{Toth:2005} and the Center for Integrated Space Weather 
Modeling (CISM) \citep[]{Luhmann:2004}.

More recently, \citet{Toth:2007} modeled
the October 28 CME as part of a Sun-to-Earth space weather event that
included an in-depth description of the resulting geomagnetic storm.
This work by \citet{Toth:2007} is ground breaking in two regards.  First,  
like those by \citet{Lugaz:2007, Cohen:2007a},  it is the first 3D numerical 
full MHD simulation to reproduce observed CMEs as they are 
magnetically driven from active regions in the low corona. 
The earliest attempt to model an observed CME (including the magnetic field)
propagating from the Sun to the Earth, is that of \citet{Wu:1999}.
This particular simulation is two-dimensional and did not capture
the structure of the solar wind or the active region, but did roughly
capture the magnetic cloud at 1 AU.  More recent modeling efforts have
been 3D and capture the structure of the solar wind, but treat CME propagation
outside of the magnetosonic point $(r \approx 20 \RS)$ \citep[]{Odstrcil:2005}.
The second noteworthy aspect of the simulation by \citet{Toth:2007} is that 
it is the first event 
study to employ a framework to couple several individual codes to model the
physical domain extending from the corona to the Earth's upper atmosphere.
The framework used was the Space Weather Modeling Framework (SWMF) 
developed by members of CSEM at the University of Michigan by 
\citet{Toth:2005}.  A framework for similar purposes is being developed
by CISM at Boston University. 

In this paper, we examine the simulation described by \citet{Toth:2007}, 
and concern ourselves with the physical properties of the
CME as it propagates from the low corona to Earth.  In particular,
the detailed structure of the coronal and solar wind allow us to quantitatively 
compare the model with Large Angle Spectrometric Coronagraph (LASCO)
C2 and C3 observations of the October 28 event.  In this case, we 
validate the accuracy of the simulation in reproducing the speed
mass, appearance (in scattered light), and shock properties of the 
observed CME.
We also make wide-angle, large elongation Thomson-Scattered white 
light images of the model to show how the CME would appear in
the wide angle coronagraphs of the Sun Earth Connection Corona 
and Heliospheric Imager (SECCHI) instrument on board the 
STEREO spacecraft.  These synthetic images show that the CME 
appears much different far from the Sun due to scattering affects 
at large elongation. 

The rest of this paper is organized as follows: The SWMF, its components,
and numerical techniques are briefly discussed in section~2.
Section~3 provides a description of the CME simulation, the results
of which are compared to LASCO observations in section~4.  
In section~5, we show how the CME would appear propagating past the
Earth in the SECCHI heliospheric imagers HI1 and HI2.
We conclude with a summary and an outlook for future development 
in section~6.

\section{Numerical Methods: SWMF and BATS-R-US}
The SWMF \citep[]{Toth:2005} couples models treating physical domains of the 
space environment extending from the solar corona to the Earth's upper 
atmosphere.  The model coupling is flexible yet efficient, making faster than 
real-time space weather simulations feasible on massively parallel computers.
Each model has its own dependent variables, a mathematical model, and a 
numerical scheme with an appropriate grid structure and temporal discretization.
The physics domains may overlap with each other or they can interact through
a boundary surface. The SWMF is able to incorporate models from the community
and couple them with modest changes in the software of an individual model.
The SWMF is a fully functional and documented framework that provides 
high-performance computational capability to simulate the physics from 
the low solar corona to the upper atmosphere of the Earth.
Currently, the SWMF is composed of nine physics modules, and is driven by  
external data such as magnetograms, flare and CME observations, satellites 
upstream of the Earth (like ACE, Geotail and Wind), etc.  
The SWMF can model any physically meaningful subset of the physics domains,
and is freely available via registration at\\
{\tt http://csem.engin.umich.edu/SWMF}.

There were seven SWMF components used in the Halloween 
storm simulations: Solar Corona (SC), Eruptive Event Generator (EE),
Inner Heliosphere (IH), Global Magnetosphere (GM), Inner Magnetosphere (IM),
Ionosphere Electrodynamics (IE), and Upper Atmosphere (UA).  For
our purposes, the only relevant components are the SC, EE and IH that
model the corona, solar wind, and CME propagation from the active
region to 1 AU.  Components SC, EE, and IH are based on the
BATS-R-US code \citep[]{Powell:1999, Gombosi:2001} that solves
the governing equations of MHD in a conservative finite volume form. 
Non-ideal MHD terms are included through appropriate source terms.
The code uses a limited reconstruction that ensures second-order accuracy
away from discontinuities, while simultaneously providing the stability
that ensures non-oscillatory solutions.
In addition, the code employs several approximate Riemann solvers.
The resulting scheme solves for the hydrodynamic and electromagnetic
effects in a tightly coupled manner 
that works equally well across several orders of magnitude
in plasma $\beta$ (the ratio of plasma pressure to magnetic pressure).
The BATS-R-US also uses a relatively simple yet effective block-based adaptive
mesh refinement (AMR) scheme to resolve structures spanning many length scales.
The BATS-R-US scales almost linearly to more than 1000 processors for a fixed
problem size, and fully implicit time stepping scheme is incorporated, 
\citep[]{Toth:2006} that can be combined with explicit time stepping. 

\begin{figure*}[ht!]
\begin{minipage}[t] {1.0\linewidth}
\begin{center}
\includegraphics[angle=0,scale=0.85]{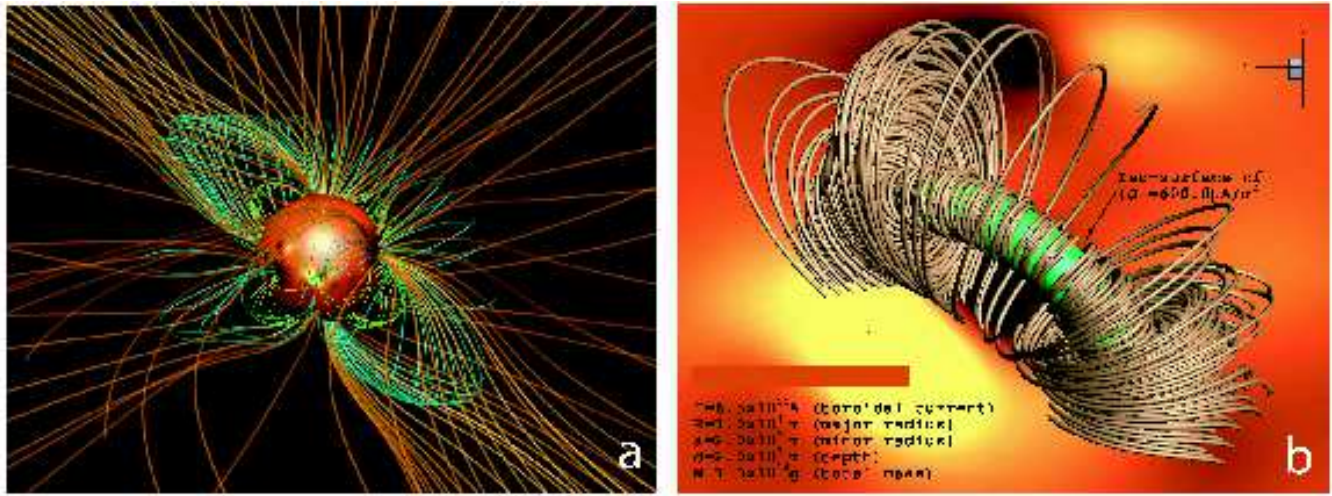}
\end{center} 
\end{minipage}\hfill
\caption{The initial condition of the corona for the October, 28 CME.
The panel on the left (a) shows the radial field strength, $B_r$ at the
base of the corona.  The structure of the coronal magnetic field is
illustrated with blue and yellow lines for the closed field while
orange lines show the open field.  The panel
on the right (b) shows a close up of AR 10486 with the superimposed
flux rope illustrated with field lines and a current density iso-surface
colored green.}
\label{initial}
\end{figure*}

The Solar Corona domain is a Cartesian box that extends from the surface 
of the Sun to $-24\,\RS < x, y, z < 24\,\RS$, where $\RS$ is the radius 
of the Sun.  The physics of this domain is described by the equations of MHD 
with additional source terms required to take into account the heating 
and acceleration of the solar wind \citep[]{Groth:2000, Usmanov:2000}. 
Recently, \citet{Cohen:2007b} developed a solar wind model based
on the empirical relationship between solar wind speed and magnetic flux tube
expansion \citep[]{Wang:1990}, which has been incorporated into the SWMF.  
Here, we use the coronal model presented in \citet{Roussev:2003b}.
At the inner boundary of the SC component, the density, pressure, 
velocity and magnetic field are defined at a height just above
the transition region. The magnetic field is obtained from a synoptic 
solar magnetogram. The boundary conditions for the
temperature and mass density at the Sun are varied as a function
of magnetic field strength to achieve a realistic distribution of
fast and slow wind speeds near the Sun and at 1AU.
The velocity components at the inner boundary maintain 
line-tying of the magnetic field to the rotating solar surface.
Differential rotation is currently neglected. 
The flow at the outer boundary is usually superfast (faster than 
the fast magnetosonic speed of the plasma), so no information
is propagating inward. 

The EE domain is in the Solar Corona, restricted to the active region 
responsible for the CME. The EE in this case takes the form of a nonlinear 
perturbation of the SC solution, which is made by superimposing a
modified version of the \citet{Titov:1999} flux rope to active region
10486.  The flux rope is in an initial state of force imbalance that
drives the eruption.  This eruption generator follows from the 
work of \citet{Roussev:2003a} who incorporated the \citet{Titov:1999} 
flux rope in the BATS-R-US code.

The IH domain extends from its inner boundary at $r=20\,\RS$ to a cube 
extending to $-240\,\RS < x,y,z < 240\,\RS$, which encompasses Earth's orbit.
The physics of this domain is described by the equations of ideal MHD, 
solved in an inertial frame, on a Cartesian grid.   
The inner boundary conditions of the IH component are obtained from the 
SC component. The flow at the outer boundary of the IH component is always 
assumed to be superfast. 

\begin{figure*}[ht!]
\begin{minipage}[t] {1.0\linewidth}
\begin{center}
\includegraphics[angle=0,scale=0.75]{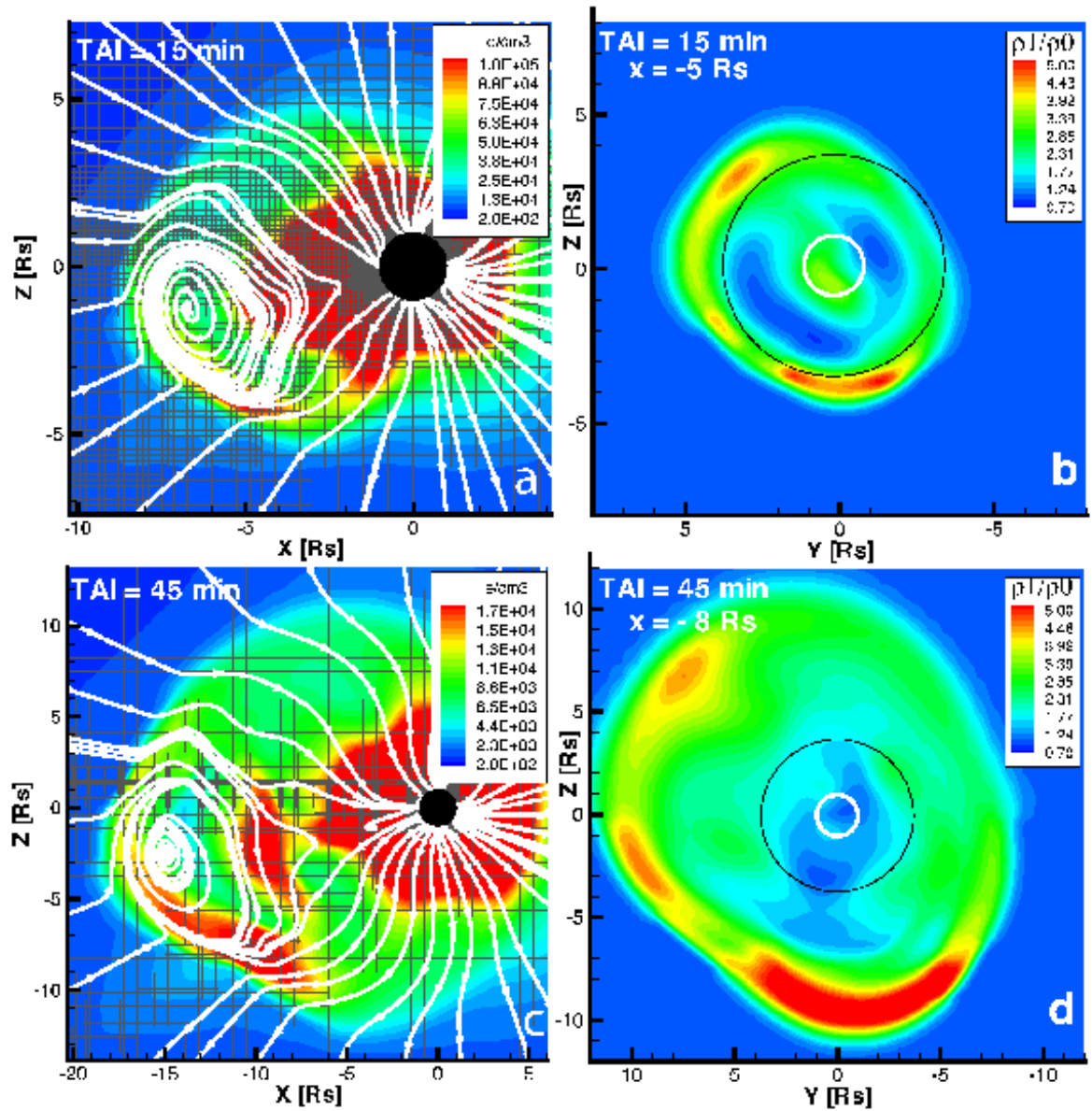}
\end{center}
\end{minipage}\hfill
\caption{The structure of the CME at time after initiation (TAI) equal
to 15 minutes and 45 minutes is shown in
the top and bottom rows respectively.  In Panels
(a) and (c) the electron density is shown in color with magnetic
stream lines (confined to the plane) drawn white and the numerical
mesh drawn black.  These images show the ejected magnetic flux
rope traveling toward the Earth in the $-x$ direction with the
center of the rope and densest plasma concentrations below the
equatorial plane.  Panels (b) and (d) show the ratio of the density
relative to the pre-event state on $y-z$ planes positioned near the
center of the CME at $x=-5 \RS$ and $x=-8 \RS$, respectively.
The white and black circles are drawn at 2 and 3.7 $\RS$, respectively,
corresponding to the occulting disks of C2 and C3, respectively.
These images show the greatest density enhancement at the bottom
$(-z)$ of a nearly circular shell type structure that evolves
in a self-similar fashion.}
\label{cmedensity}
\end{figure*}

\section{Halloween Storm Simulation}
\subsection{Initial State}
Here, we summarize the basic description of the CME simulation
that was first presented in \citet{Toth:2007}.  For more details,
we refer the reader to this earlier paper.  The simulation begins
with the construction of a steady state corona and solar wind.
The magnetic field at the inner boundary (base of the corona) of the 
SC component is specified from an MDI synoptic magnetogram centered 
around the time of the October 28 eruption.  The initial volumetric
field is specified with a potential-field-source-surface 
extrapolation fit to this map. The computational grid is highly
refined around AR 10486 where the smallest cells are about 
$3\times10^{-3}\,\RS$.  The grid is refined to $0.1\,\RS$ within a
$1\,\RS$ diameter cylinder  extending along the Sun-Earth line.
The SC component is then run to steady state allowing the 
solar wind to relax.
Figure~\ref{initial}, Panel (a) shows the steady state solution. 
Here the radial field strength is shown in color at the inner boundary, 
and field lines are drawn extending out into the corona. 
Orange lines are open while closed lines are shown in blue. 
The SC is run in the Heliographic Rotating (HGR) coordinate
system with the sidereal Carrington rotation period of 25.38 days.
An offset angle around the $Z$ axis places the Earth in the 
$-X,Z$ half plane at the time of the large eruption on October 28 CME. 

The IH component is coupled to the SC component at $20\,\RS$, and rapidly 
achieves steady state as the superfast wind blows through the domain. 
During the steady state run, the IH component uses a rotating coordinate 
system, which is then switched to Heliographic Inertial (HGI) with
an offset angle that puts the Earth in the $-X,Z$ half plane
(the orbital motion of the Earth is neglected).  The grid resolution
is $1/4\,\RS$ near the inner boundary and also along the Sun-Earth  
line within a cylinder of $1\,\RS$ radius, while the largest cells 
are $4\,\RS$. 

\subsection{Generating Eruptive Events}
The October 28 event was preceded by several smaller CMEs, which 
significantly modified the ambient solar wind. 
To take into account the preconditioning effect of previous CMEs on the solar 
wind, we start our simulation on October 26, when a smaller CME was launched 
at approximately 07:00 UT from AR 10486.  For this first CME, we use 07:24 UT 
as the initiation time, which is within the uncertainties of the observations.
The much more energetic CME that is the focus of this study occurred at
approximately 10:54 UT.  This eruption occurred near disk center
producing a full halo CME.

\begin{figure*}[ht!]
\begin{minipage}[t] {1.0\linewidth}
\begin{center}
\includegraphics[angle=0,scale=0.85]{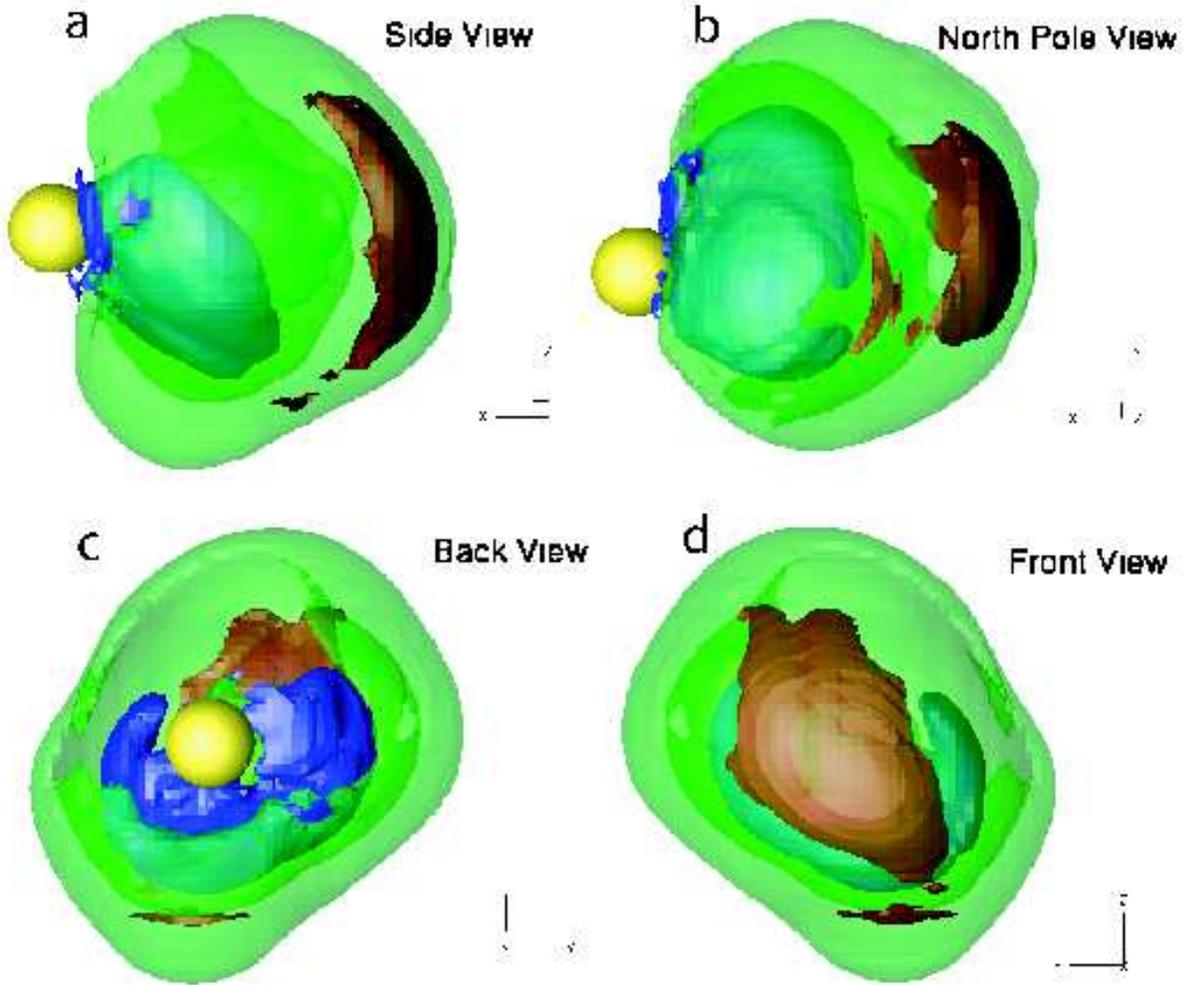}
\end{center}
\end{minipage}\hfill
\caption{Three-dimensional density structure of the model CME 15 minutes after
initiation.  Iso-surfaces of the density enhancement (relative to background)
are shown at levels of 0.9, 2 and 5 in blue, green and red respectively.
The green iso-surface shows the extent of the shock front, while the blue
iso-surface shows the low density rarefaction behind the CME. The Sun is
shown with a yellow sphere.}
\label{3dCME1}
\end{figure*}

\begin{figure*}[ht!]
\begin{minipage}[t] {1.0\linewidth}
\begin{center}
\includegraphics[angle=0,scale=0.90]{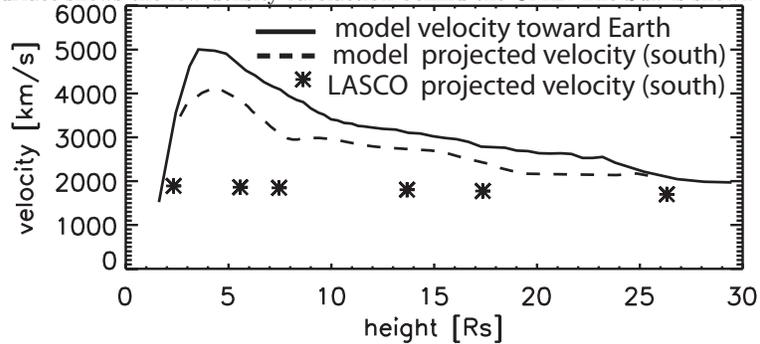}
\end{center}
\end{minipage}\hfill
\caption{Comparison of observed and modeled CME velocity.
The solid line shows the modeled CME velocity moving
directly toward the Earth, while the dashed line shows the model
velocity projected on the plane of the sky 177 degrees (counter clock wise)
from the north polar axis.  At this same location in the plane of the sky,
the CME velocity is derived from LASCO observations and plotted with stars.
We find that the model briefly reaches a velocity of 4000 km s$^{-1}$ at
$4.5 \RS$ before falling to 2000 km s$^{-1}$ at $20 \RS$.
In contrast, the CME is observed to decelerate from 1890 km s$^{-1}$
to 1699 km s$^{-1}$ as it travels from $2.3 \RS$ to $26.3 \RS$.}
\label{velocity}
\end{figure*}

\begin{figure*}[ht!]
\begin{minipage}[t] {1.0\linewidth}
\begin{center}
\includegraphics[angle=0,scale=0.60]{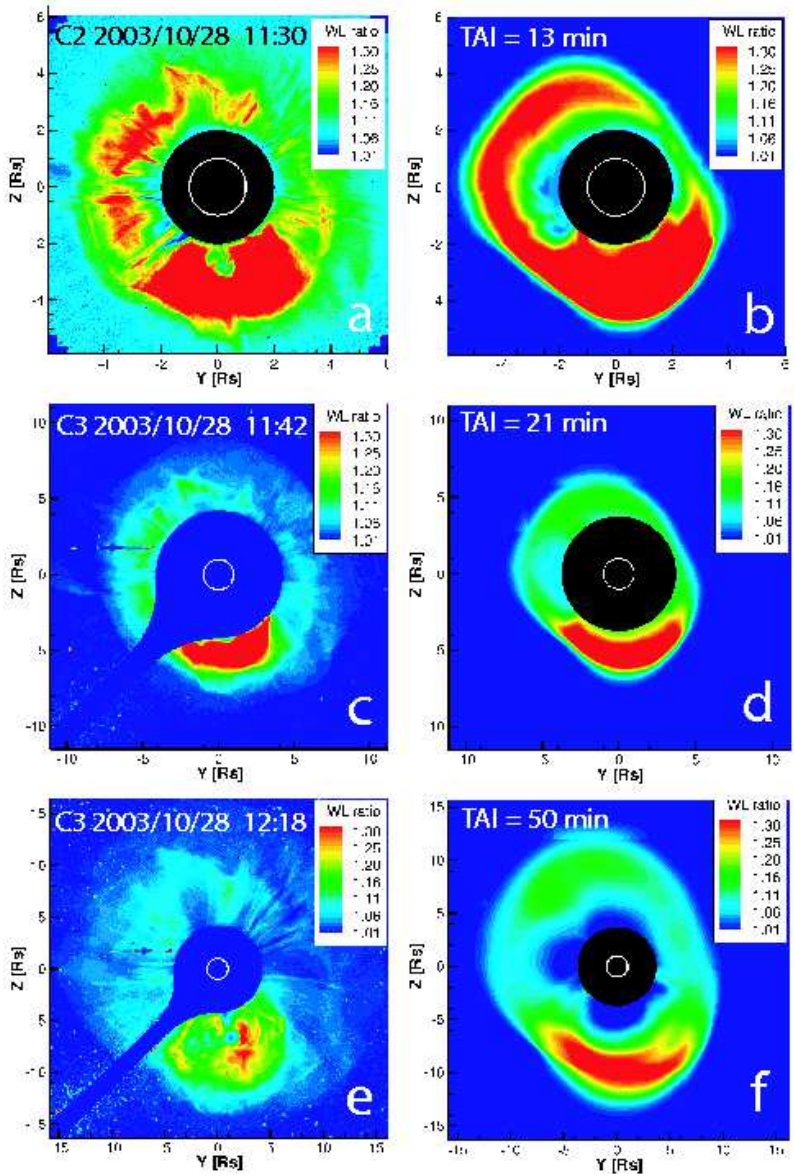}
\end{center}
\end{minipage}\hfill
\caption{Comparison of observed (left) and simulated (right)
Thomson-scattered white light brightness.  The color images show
the total brightness divided by that of the pre-event background.
White circles show the solar limb, filled black circles show occulting disks.
Panel (a) shows the LASCO C2 observation at 11:30 UT, while LASCO
C3 observations are shown in panels (c) and (e) at 11:42 and 12:18,
respectively.  Panels (b) (d) and (f) show model coronagraph images
at 13, 21 and 50 minutes after initiation.  Here,  we find that the model
achieves remarkable quantitative agreement with both the magnitude and
spatial distribution of the observed brightness.}
\label{lascomparo}
\end{figure*}

\begin{figure*}[ht!]
\begin{minipage}[t] {1.0\linewidth}
\begin{center}
\includegraphics[angle=0,scale=0.90]{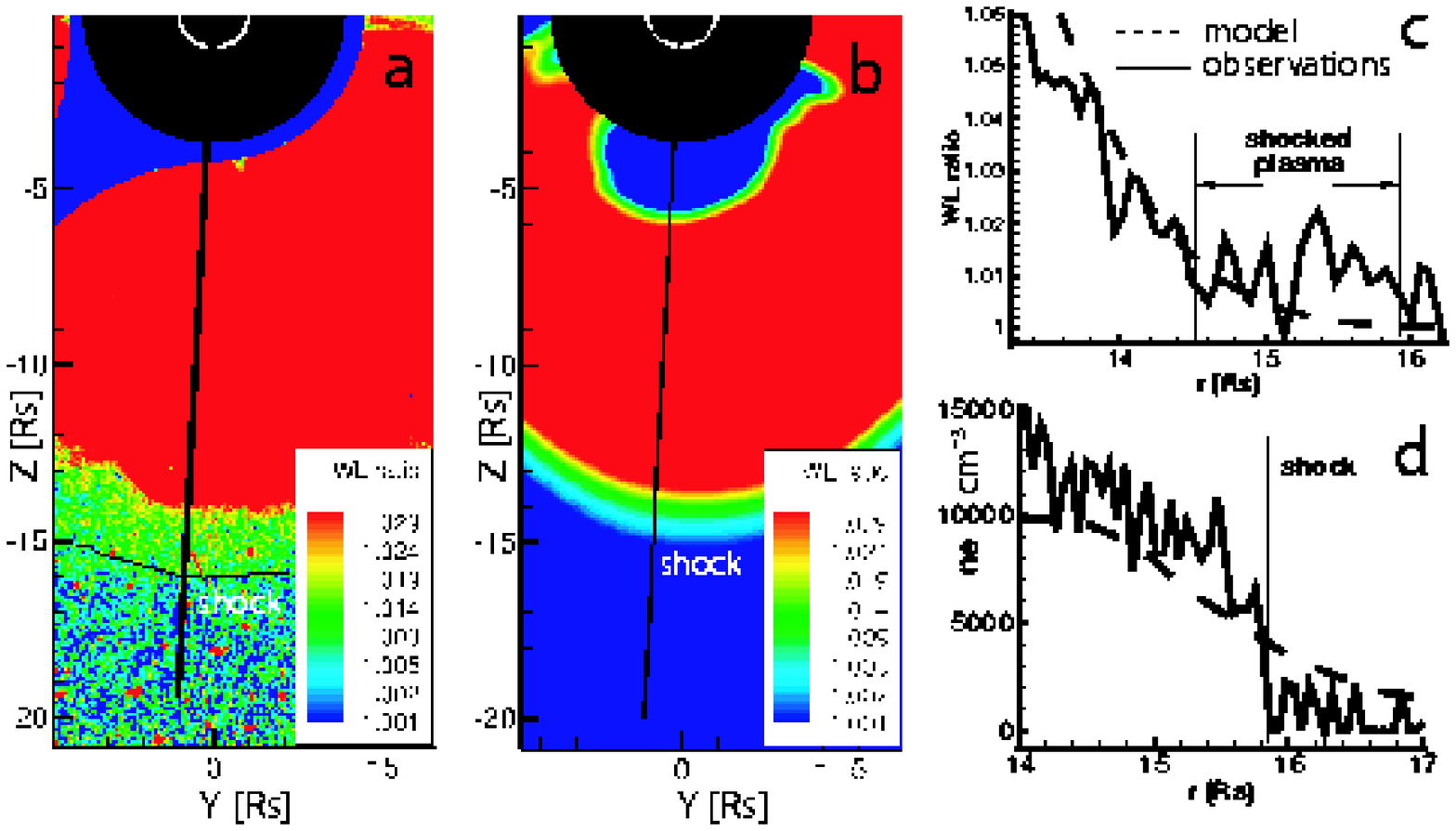}
\end{center}
\end{minipage}\hfill
\caption{Shock structure as revealed by coronagraph images.
Panels (a) and (b) respectively show color images of the total brightness
(divided by the pre-event brightness) for LASCO C3 (time = 12:18)
and for the simulation  50 minutes after initiation.  The color legend is
chosen to highlight the faint (2 \%) increase in brightness
at the shock found at the outer edge of the halo.  The model
shows the same faint increase in brightness at the shock as well as the
capturing the width of the faint plasma sheath.  Panels (c) and (d) show
respectively line plots of the brightness and electron density as
functions of distance from Sun center along the black line at 177
degrees from the north polar axis.  Model and observations are
shown with dashed and solid lines respectively.  The line plots
reveal that the simulation quantitatively reproduces the observed
brightness increase behind the shock as well as the inferred density
increase at the shock.}
\label{shock}
\end{figure*}

Here, we initiate the CMEs by inserting magnetic flux 
ropes (the size of the active region) based on the modified 
\citet{Titov:1999} model.  In Figure~\ref{initial} (Panel (b)), the
flux rope for the October 28 eruption is seen arching above AR 10486.   
The location and orientation of the flux ropes 
were chosen to arch over the dipolar part of the active region.
The active region can be easily identified in the high resolution synoptic 
map obtained from the observed photospheric magnetic field.  
The density in the loop was obtained from the size and estimated mass
of the CMEs.  The rope's
magnetic field does not match that observed at the photosphere.  The reason 
for the discrepancy is that the field of the rope is highly twisted 
and is orthogonal to the polarity inversion line. 
In contrast, the measured magnetic field was observed
to be in a highly sheared configuration running
parallel to the inversion line \citep[]{Liu:2005}.
In spite of this limitation, the magnetic field strength 
of the flux ropes can be adjusted to produce eruptions that
match LASCO observations of the CME speeds at 20 solar radii
(1500 and 2500\,km s$^{-1}$ for the October 26 and October 28 events, 
respectively).  For the first CME, the magnetic flux rope has a 
free enegy $2.3\times 10^{32}$ ergs, while the second faster 
CME requires nearly ten times as much energy at $2.0\times 10^{33}$ ergs.

After 20 minutes from the start of the time dependent simulation, the 
leading shock of the first CME reaches a radial distance of 5.5$\,\RS$ 
at a speed slightly exceeding 2100\,km s$^{-1}$.  
The first CME reaches the SC/IH boundary at 
20$\,\RS$ after about 1.7 hours, and the speed of the leading shock is 
the observed 1500\,km s$^{-1}$. The first CME is propagating in a direction 
about 30 degrees off from the Sun-Earth line, but the flanks of the shock 
reach the Earth 45 hours after the eruption at 4:30 UT Oct 28. The simulated  
solar wind velocity at the Earth increases from about 350 km s$^{-1}$ 
to 550 km s$^{-1}$, 
which is a good approximation to the solar wind conditions preceding 
the arrival of the October 28 CME as observed by the ACE satellite
\citep[]{Skoug:2004, Zurbuchen:2004}.   

The second CME is initiated at 10:54 October 28 with a flux rope that is 50\%
larger in radius than the previous one.  Due to the larger size and stronger 
magnetic field, the second CME reaches a speed of 2500\,km s$^{-1}$ (on the  
Sun-Earth line) measured at 20$\,\RS$.  After 15 minutes, the second CME  
reaches 9$\,\RS$ with a shock speed near 3200\,km s$^{-1}$.  The shock reaches 
the SC/IH boundary at 20$\,\RS$ in less than an hour with a speed around 
2800\,km s$^{-1}$.  The density structure of the October 28 CME is shown in
Figure \ref{cmedensity}.  Panels (a) and (c)  show a color 
representation of the electron density on the meridional plane $(x-z)$ at 15
and 45 minutes after CME initiation, respectively.  White lines show 
the direction of the magnetic field confined to the plane.  
In these pictures, the magnetic flux rope is clearly seen to be
expelled from the corona.  Panels (b) and (d) show the density ratio
(relative to the pre-event corona) on $y-z$ located at $x = -5 \RS$
and $x = -8 \RS$ at 15 and 45 minutes after initiation, respectively.
The white and black circles correspond to the solar limb and LASCO
C3 coronagraph, respectively.  These images show the nearly self-similar
evolution of the plasma expelled in the CME.  The plasma is 
distributed in a shell with the greatest density in the southern
hemisphere.

The three-dimensional density of the CME is shown in Figure~\ref{3dCME1} 
with iso-surfaces of the density enhancement (i.e. density divided by
the pre-event level) at $t= 15$ minutes after initiation. 
Blue, green and red iso-surfaces corresponding to values of 0.9, 2, and 5
are shown from four perspectives: (a) side, (b) polar, (c) back and 
(d) front facing Earth.  The green surface effectively shows the extent
of the shock front, the red surface shows the core material, while the blue 
iso-surface shows the density depleted cavity.  The cavity forms as 
a rarefaction behind the flux rope, while the densest plasma is located 
in the southern hemisphere, both in and below the flux rope.  The 
polar and back views (Panels (b) and (d) respectively) best show the
density enhancement at the southern end of the CME.
Comparison of Figure~\ref{3dCME1} with Figure~6 in \citet{Jackson:2006}
also suggests that the simulated 3D density structure qualitatively agrees
with the density enhancements reconstructed from SMEI observations.

\section{Comparisons with LASCO Observations}
CMEs are most frequently observed in visible light that is Thomson 
scattered by coronal electrons within the CME.  In addition to this electron
scattered photospheric light (referred to as the K-corona), there is also 
light scattered by interplanetary dust that is referred to as the F-corona.  
There are significant differences between the two components of scattered 
light.  First, the F-corona is unpolarized within 5-6$\,\RS$, while the 
K-corona is polarized with components both radial and tangential to the solar 
limb \citep[]{Billings:1966}.  Second, the brightness of the K-corona falls  
off much more rapidly with distance from the Sun such that it dominates close
to the Sun, is roughly equal to the F-corona at $r \approx 4\,\RS$, 
and is much dimmer than the F-corona far from the Sun.  Finally, the
F-corona is nearly constant in time and is believed to be largely 
unaffected by CMEs. 

The corona is optically thin in white light,
so that the intensity of scattered light along a 
given line of sight (LOS) is the integrated contribution of all
electrons along the line.  The contribution to a given LOS is 
highly dependent on the location of the electron relative to the
Sun and the observer \citep[]{Billings:1966}.  Close to the Sun, the 
scattering is heavily weighted in the plane of the sky.  Coronagraphs
located in space observe coronal light without contending with
atmospheric scattering, which makes it possible to view the
corona far from the Sun.  In the case of LASCO, the C2 and C3 
coronagraphs have fields of view extending to 6 and 32$\,\RS$
respectively.  More recently, SMEI and STEREO coronagraphs have 
observed Thomson scattered light at very large angles from the Sun 
that extend beyond Earth's orbit.

In this section, we make comparisons between synthetic
coronagraph images constructed from our model CME, and LASCO observations.
The synthetic images are created by numerically integrating the Thomson
scattered light along a LOS for each pixel with the appropriate
scattering function \citep[]{Billings:1966}.  Images of numerical
CME models have been made in this fashion before  
\citep[e.g.]{Wu:1999, Odstrcil:2003, Manchester:2004a, Lugaz:2005}.  More 
recently, synthetic images have been made for models of specific CME events, and
qualitatively compared to LASCO images \citep[e.g.][]{Lugaz:2007, Cohen:2007a}.
Here, we go further, and for the first time make truly
quantitative comparisons with the data.  To make the comparison, we 
process the synthetic data and the LASCO data in the same way and
display the resulting images in identical formats.  We account
for F-corona far from the Sun where it is the dominant source
of scattered light.  Since our MHD model only allows us to directly treat
electron density, we estimate the contribution of dust scattering 
with the same power-law empirical relationship used to make the
LASCO images.  In this case, the background F-corona brightness
is taken to have the form $B_F = c r^{(0.22\cos(2\theta) - 2.47)}$
\citep[]{Koutchmy:1985}. 
Here, we estimate the magnitude of $c$ by setting $B_F = B_K$ at
$r = 4\,\RS$ where $B_K$ is the background polar brightness
of the K-corona taken from \citet{Saito:1977}.  

\begin{figure*}[ht!]
\begin{minipage}[t] {1.0\linewidth}
\begin{center}
\includegraphics[angle=0,scale=0.90]{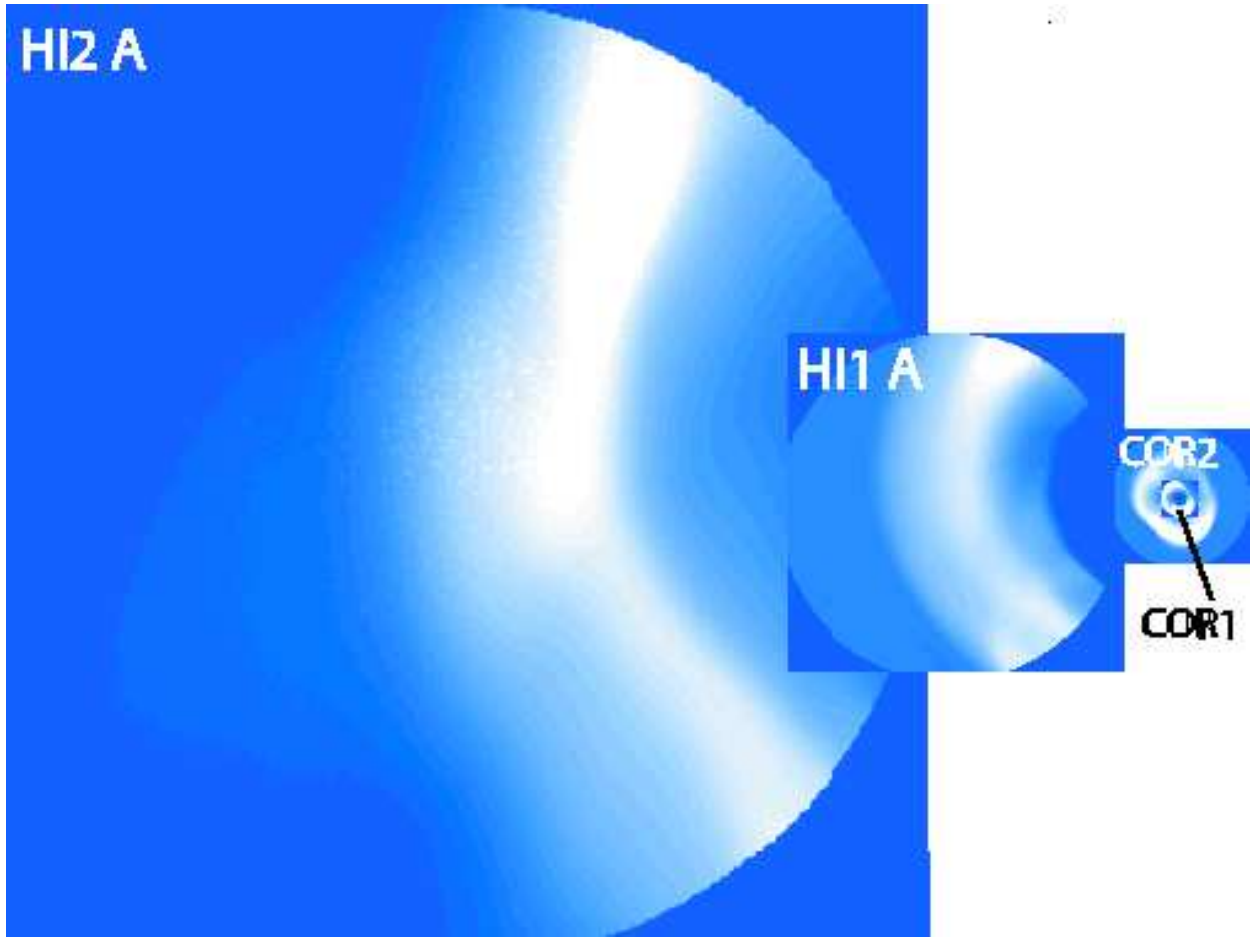}
\end{center}
\end{minipage}\hfill
\caption{Views of the model CME as it would appear in the
SECCHI coronagraphs.  The CME is shown in a time series
propagating through the fields of view of COR1, COR2, HI1 and HI2
that extend from 1.4 to 4$\,\RS$, 2 to 15$\,\RS$, 12 to 84$\,\RS$
(20 degrees wide), and 66 to 381$\,\RS$ (70 degrees wide) respectively.
In this case, the HI images are
taken from the point of view of STEREO A, 22 degrees ahead of the
Earth, while COR images are made from the location of Earth.}
\label{STEREO}
\end{figure*}

The results of the comparison are shown in Figure~\ref{lascomparo}.
Here, images of the October 28 CME as observed with LASCO C2 (Panel (a)) 
and C3 (Panels (c) and (e)) are shown for times $t =$ 11:30, 11:42
and 12:18 respectively.  Panels (b), (d), and (f) of Figure~\ref{lascomparo}
show the synthetic images at times $t =$ 13, 21, and 50 minutes after
initiation when the model CME is closest in size to that of the corresponding
LASCO image.  In reality, the CME travels slower than in our model
with the observed sequence of images occurring at approximately 36, 48, and 84
minutes after initiation.  In both data and model, the color images show
the total brightness divided by that of the pre-event background.
White circles show the solar limb, and filled black circles show the occulting 
disks.  In all cases, we find extremely good quantitative agreement 
in both the magnitude and spatial distribution of the observed brightness.  
In these images, the greatest brightness is below the southern limb of
the Sun, that extends in an arc over the eastern limb, while the 
the dimmest part of the CME is found in the north western 
quadrant. This brightness distribution corresponds with the position of 
AR 10486, which was located at 20 degrees below the equator on October 28. 
By 12:18 UT, the observed brightness patterns is 
more highly structured than the model, showing evidence
of a bright core in the south.  The model (Panel (f)) lacks these
fine details, and retains a single bright arc in the south. 

\subsection{CME Mass}
To further quantify the comparison between the our numerical model and
the LASCO coronagraph observations, we calculate the CME mass in both
cases in an identical fashion.  This mass is derived by integrating 
the excess brightness of the K-corona  with the assumption 
that the plasma is in the plane of the sky, which sets a lower 
limit on CME mass.  The F-corona is subtracted out from the observations
and does not affect the mass.  We restrict this integral to the brightest 
portion of the CME that extends from 235 to 330 degrees as measured
from the $-y$ axis in Figure \ref{lascomparo}.  In the case of
the C2 field of view (top row of Figure \ref{lascomparo}), this
integration sector extends from 2.0 to 6.0 $\,\RS$, and for the C3 
field of view (middle and bottom rows of Figure \ref{lascomparo}) 
the sector extends from 4.0 to 16.4$\,\RS$.  With these criteria,
the CME masses derived from observations at $t =$ 11:30, 11:42 and
12:18 UT are $1.50 \times 10^{16}$ g, $1.55 \times 10^{16}$ g, and 
$1.70 \times 10^{16}$ g, respectively.  The corresponding masses of 
the model CME (Panels (b), (d), and (f)) are,
respectively, $2.23 \times 10^{16}$ g, $2.16 \times 10^{16}$ g, and 
$4.90 \times 10^{16}$ g.

The observed masses are approximately 30\% less than that of the model at times
$t =$ 11:30 UT and 11:42 UT, which is consistent with the larger filling
factor of the model event. 
The slight decrease in model CME mass found from Panel (d) is the 
result of a greater part of the CME being obscured by the occulting disk.
By 12:18 the CME mass is observed to increase by 13 \%, while the 
model mass (as shown in Panel (f)) more than doubles.   
We expect the mass of fast CMEs to increase as plasma is swept up as  
they travel from the Sun as shown in numerical simulations 
\citep[]{Manchester:2004a, Lugaz:2005}.  However, the mass increase seen in
LASCO observations is attributed to mass entering the coronagraph
field of view from below the occulter \citep[]{Vourlidas:2000}.  In general,
there is not yet proof of a plasma pileup in the low corona by CMEs.  
In the case of our simulation, we believe that the CME mass increases
faster than what is observed because of the excess speed of the CME.
For the observed event, the (excess) mass continues to increase by
almost an order of magnitude, to $13.6 \times 10^{16}$ g as measured 
with SMEI on October 29, 12:00 UT as the CME  passed the Earth 
\citep[]{Jackson:2006}.

\begin{figure*}[ht!]
\begin{minipage}[t] {1.0\linewidth}
\begin{center}
\includegraphics[angle=0,scale=0.90]{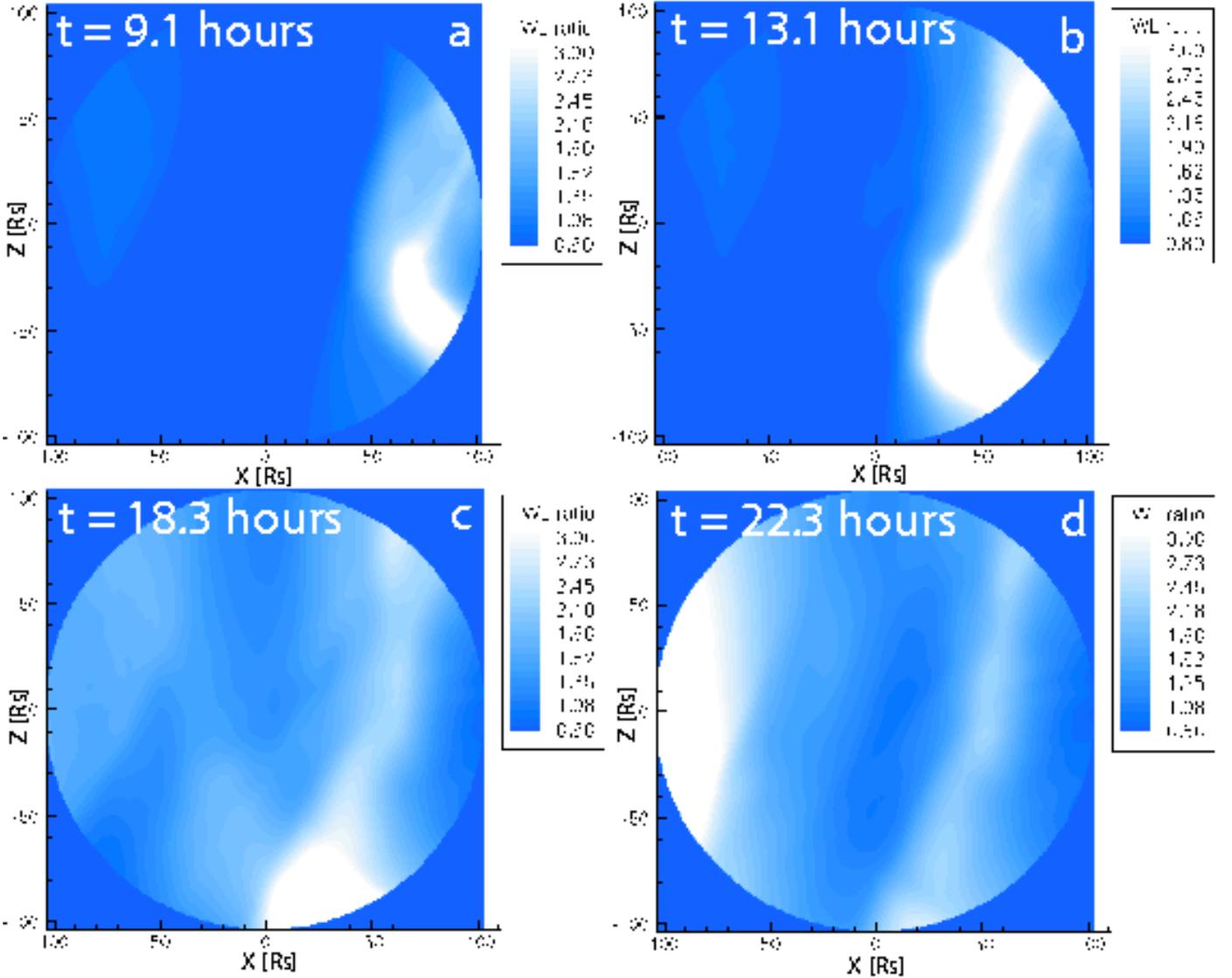}
\end{center}
\end{minipage}\hfill
\caption{Synthetic image of the CME as it would appear from
SECCHI-HI2A one year into the mission (22 degrees ahead of the
Earth).  The total brightness is shown in a time sequence
at intervals $t =$, 9.1, 13.1, 18.3, and 22.3 hours in panels
(a), (b), (c) and (d), respectively. Here, we see the CME propagate
into the field of view on the right hand side of Panel (a), appear
to stall in Panel (b), and then begin to brighten on the left hand
sides of Panels (c) and (d).  This type of evolution is not observed
close to the Sun, and occurs at large elongation because of the
passage of the CME density structure over the spherical surface
of maximum scattering.}
\label{HI2A}
\end{figure*}

\subsection{Shock Identification} 
Faint arcs are frequently observed at the outer edges of 
fast CMEs viewed in coronagraph images, yet it has been difficult to
conclusively identify these arcs as shocks.   
The earliest example of shock identification is that of \citet{Sime:1987} 
who observed a bright loop at the front of a fast CME. 
In that case, the presence of a shock was inferred from its high speed
(1070 km s$^{-1}$), the absence of any deflections preceding the shock,
and the fact that the expanding shock front did not cease
its lateral motion to form stationary legs.  More recently,
\citet{Vourlidas:2003} observed a similar feature with LASCO that they
identified as a shock wave.  In this work, a numerical simulation 
was employed to model the shock's appearance and confirm its presence in 
the coronagraph image.  More recent MHD simulations have also found 
manifestations of CME driven shocks in synthetic coronagraph 
images \citep[e.g.][]{Manchester:2004a, Lugaz:2007}.  
Perhaps the most compelling observational evidence for shocks appearing  
as visible components of CMEs in LASCO images is presented in 
\citet{Raymond:2000} and \citet{Mancuso:2002}.  Here, shocks were 
observed simultaneously in the low corona $(r < 3\,\RS)$ by LASCO, the
Ultraviolet Coronagraph Spectrometer (UVCS) and as type II radio
burst.  UVCS gave clear spectroscopic evidence for the presence
of shock fronts, while radio bursts indicated the presence of 
shock-accelerated electrons.  

\begin{figure*}[ht!]
\begin{minipage}[t] {1.0\linewidth}
\begin{center}
\includegraphics[angle=0,scale=0.90]{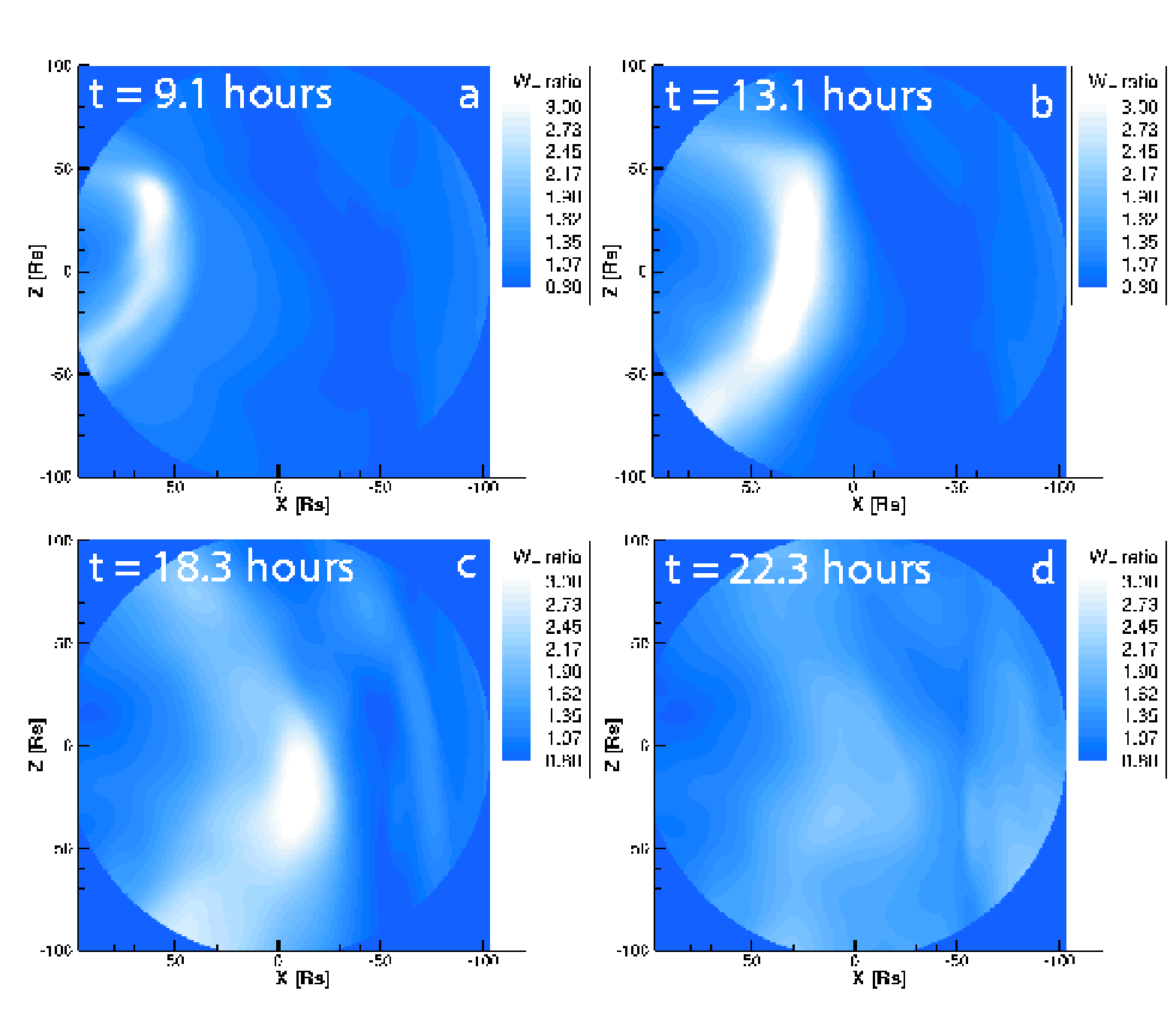}
\end{center}
\end{minipage}\hfill
\caption{Synthetic image of the CME as it would appear from
SECCHI-HI2B one year into the mission (22 degrees behind the
Earth).  The total brightness is shown in a time sequence
at intervals $t =$, 9.1, 13.1, 18.3, and 22.3 hours in panels
(a), (b), (c) and (d), respectively. Here, we see the CME propagate
into the field of view on the left hand side of Panel (a), noticably
brighten in Panel (b), fade in Panel (c), and then all but disappear
in Panel (d).  This evolution is much different than that shown
in Figure \ref{HI2A}, and illustrates the 3D structure viewed
in the opposite direction as seen by the HI2A coronagraph.}
\label{HI2B}
\end{figure*}

\begin{figure*}[ht!]
\begin{minipage}[t] {1.0\linewidth}
\begin{center}
\includegraphics[angle=0,scale=0.90]{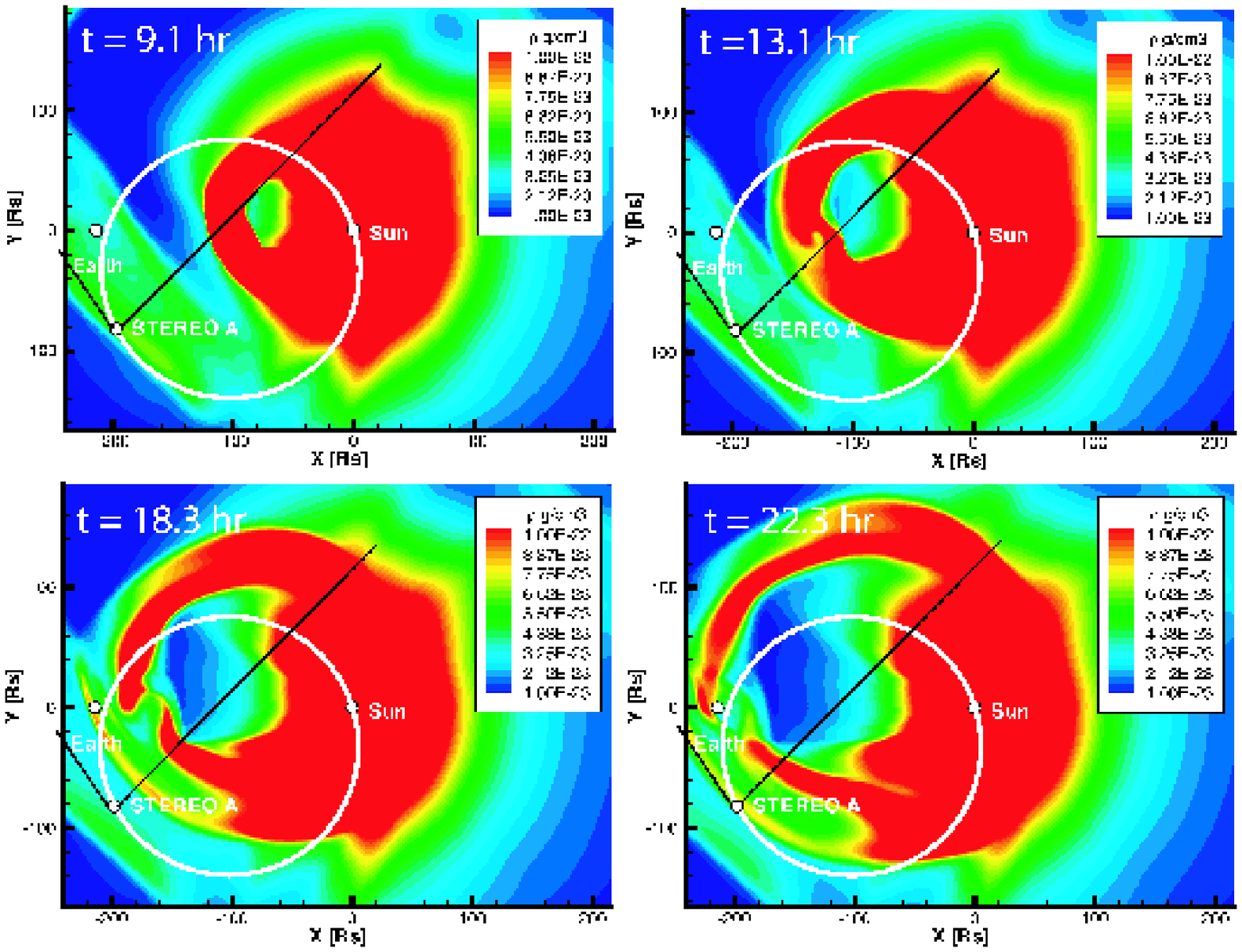}
\end{center}
\end{minipage}\hfill
\caption{The time evolution of the density structure of the CME on the
equatorial plane.  The density is shown in color in a time sequence
at intervals $t =$, 9.1, 13.1, 18.3, and 22.3 hours in panels (a), (b), (c) and
(d) respectively. The locations of the Sun, Earth, and STEREO A and it's
field of view are shown along with the equator of the Thomson sphere,
the location of maximum scattering into the observer's LOS.  The
panels correspond to the same times shown in Figure \ref{HI2A}, and
clearly show how dense plasma crosses the Thomson sphere to produce
the time evolution of brightness seen in Figure \ref{HI2A}}.
\label{rho1AU}
\end{figure*}

In our simulation, we can conclusively identify a shock front and 
determine its appearance in synthetic coronagraph images. 
By detailed comparison with the observations, we can then clearly 
demonstrate the existence of a shock front in the corona.
For this purpose, we reexamine the coronagraph data presented in 
the bottom row of Figure \ref{lascomparo}.  We again show the brightness, 
but with the color contours adjusted in Figure \ref{shock} to more clearly 
show the faint features at the outer-most edge of the observed and
modeled CME in Panels (a) and (b) respectively.  
In both panels, we see very similar faint rims at the outer 
edge of the CME where the color makes a transition from
blue-green and yellow indicating an increase in brightness at
a level between 1 and 2\%.  This rim has a width of 2$\,\RS$ and extends  
out to $16\,\RS$ from the Sun.  In the data, this faint rim is near the 
noise level and we draw a curved black line to mark its outer boundary.

For a more detailed comparison, we extract the brightness and electron
density along radial lines (shown in Figure \ref{shock}, Panels (a) and (b))
that pass through the southern outer edge of the CME.  
Panel (c) shows both the observed and modeled brightness with solid
and dashed lines respectively.  We find very good agreement between
the model and the observations.  Both cases exhibit a similar steep 
drop in brightness that levels off forming the faint rim, where
the noise in the observations is close to the signal level. 
For the purpose of the comparison, we have aligned these features in 
the radial direction by shifting model results a small amount $(+0.62\,\RS)$.
In Panel (d), we plot (solid line) the electron density deduced from 
observations under the assumption of an axisymmetric distribution 
around the Sun.  The model electron density is extracted along
a line extending from the center of the Sun, and passing through the 
southern most extent of the disturbance near $x,y,z = -8.6, 0.83, -15.9\,\RS$. 
The projection of this line on the $y-z$ plane corresponds to line 
shown in Panel (b).  We plot this data in Panel (d) with a 
dashed line and translate the model density in the radial
direction to align with the observed quantities.   
We find remarkable agreement between the two electron densities, 
where both increase from approximately 2000 to 10000 cm$^{-3}$ 
at the outer edge of the faint rim between 14 and $17\,\RS$.
In the simulation, we can conclusively identify this density jump
as a strong fast-mode shock driven by plasma moving at 2010 km s$^{-1}$.
The density increase at the shock is very near the theoretical limit of
$(\gamma +1)/(\gamma -1) = 5$ for $\gamma = 1.5$ used in the simulation.
The shock is smeared out over a distance of $2\,\RS$ by grid 
resolution at $1/4\,\RS$. By a detailed comparison between the simulation  
and the observations, we clearly demonstrate the existence of a shock front 
in the LASCO C3 image at $t =$ 12:18.

The overall 3D shape of the shock front is seen in Figure \ref{3dCME1}, 
from which we can discern why the shocked plasma is so faint. 
The outer edge of the shock front is $5\,\RS$ from the plane of 
the sky and is sharply curved at its greatest distance from the
$x$ axis.  The lines-of-sight at the outer edge of the CME graze the
shock surface, and even though there is nearly a factor of four 
increase in density at shock, the enhancement occurs over such a
short distance that it results in a very small increase in brightness.
The brightness increases as the lines of sight pass through 
progressively longer distances of compressed plasma behind the shock. 
For the numerical model, the decrease in brightness occurs faster
than what is observed, in part because the shock front is smeared
out by the limited numerical resolution.  

\section{CME Appearance at Large Elongation: STEREO-SECCHI Predictions}

In this section, we make wide-angle, large elongation Thomson-scattered white
light images of the model to show how the CME would appear in the 
coronagraphs of the Sun Earth Connection Corona and Heliospheric Imager
(SECCHI) instruments onboard the STEREO spacecraft \citep[]{Howard:2007}.
The fields of view of the four SECCHI 
coronagraphs: COR1, COR2, HI1, and HI2 give complete coverage extending 
from the low corona to beyond the Earth's orbit as shown in Figure 
\ref{STEREO}.  COR1, COR2 coronagraphs are centered on
the Sun with fields of view that extend from 1.4 to 4$\,\RS$ and 
2 to 15 $\RS$ respectively.  HI1 and HI2 are pointed toward the 
Sun-to-Earth line with fields of view that extend from 12 to 84$\,\RS$ 
and 66 to $381\,\RS$, respectively.  Figure \ref{STEREO} shows
synthetic COR1 and COR2 images of our model of the October 28
CME as it appears from the location of the Earth.  HI1 and HI2
images show the CME from a location 22 degrees ahead of the Earth where
STEREO A will be one year into it's mission.

Thomson scattering has a strong angular dependence such that scattering
into a LOS is highly dependent the on location of the electron relative to the
Sun and the observer \citep[]{Billings:1966}.  The region of space for which 
there is maximum scattering into LOS is in the shape of a sphere that contains
the Sun and the observer as see in Figure \ref{rho1AU}.
Close to the Sun, the scattering is heavily weighted to the plane
of the sky.  At large elongation (more then $60\,\RS$ from the Sun), 
the spherical shape of the maximum scattering surface becomes
significant to the appearance of coronagraph images \citep[]{Vourlidas:2006}.

Our model provides an opportunity to examine the effects of the angular
dependence of Thomson scattering on the appearance of CMEs at large 
elongation as seen in the HI2 coronagraphs. 
Figure \ref{HI2A} shows the time evolution of the model CME as it would appear 
from SECCHI-HI2A one year into the mission (22 degrees ahead of the
Earth).  The total brightness is shown in a time sequence
at intervals $t =$, 9.1, 13.1, 18.3, and 22.3 hours in panels
(a), (b), (c) and (d), respectively. The CME is seen to propagate
into the field of view on the right hand side of Panel (a), appears
to stall in Panel (b), and then reappears on the left hand
sides of Panels (c) and (d) without ever being bright in the middle
of the field of view.  (A movie is available for the online version 
of this paper.)  Figure \ref{HI2B} shows the corresponding 
series of images that would be made be SECCHI-HI2B.  In this case,
the spacecraft is 22 degrees behind the Earth with the HI2B instrument
pointing toward the Sun-Earth line in the opposite direction as HI2A.
From this perspective, the CME enters the HI2B field of view from the 
left hand side seen in Figure \ref{HI2B}, Panel (a). Here, the CME appears
as an arc that first brightens as seen in Panel (b) and then gradually
fades away as seen in Panels (c) and (d). 

The complex behavior shown in the model HI2 images is not seen in 
standard coronagraph images, and was not found in the simulated 
HI2 images of a less structured CME examined by \citet{Lugaz:2005}.
The reason the CME fades in and out is because of multiple crossings of 
the Thomson sphere by dense arcs that propagate through the field of view
of HI2A as illustrated in Figure \ref{rho1AU} \citep[]{Vourlidas:2006}.
Here, for the same time sequence as shown in Figure \ref{HI2A},
we show the mass density in color on the equatorial ($x-y$)
plane.  We mark the intersection of the Thomson sphere with the
equatorial plane with a white circle.  The position of STEREO A 
is marked along with black lines roughly showing the field of
view of the HI2 coronagraph.  Dense plasma is preferentially brightened
in close proximity to the Thomson sphere.  At $t = 9.1$ hours, a dense
arc of plasma crosses the Thomson sphere from the right hand side.  
This arc begins to move away from the Thomson sphere followed
by a cavity at times $t = 13.1$ and 18.3 hours, which explains the
dimming on the right hand side of the field of view.  By time 
$t = 22.3$ hours, dense plasma on the Thomson sphere is found 
directly in front of the space craft on the left hand side, 
and far across the Thomson sphere on the right hand side.  
This two-sided distribution of plasma and its proximity
to the Thomson sphere coincides with the bright arcs seen in 
the left and right hand sides of the field of view seen in 
Panel (d) of Figure \ref{HI2A}.

\section{Summary and Conclusions}

We have preformed a detailed analysis of a 3D MHD time-dependent 
simulation of flux-rope-driven CMEs originating from AR 10408 on October
26 and 28, 2003 \citep[]{Toth:2007}.
The simulation was carried out with the SWMF in which propagation of
the CMEs were followed from the low corona to beyond 1 AU.
We have created synthetic coronagraph images of the October 28 event
and made strict quantitative comparisons of these images with those
made my LASCO C2 and C3.  The simulation closely reproduces the 
appearance of the October 28 CME, both in the morphology and magnitude of 
Thomson scattered light through the fields of view of both LASCO C2 and C3.
Furthermore, the CME masses calculated from the synthetic coronagraph images
and the LASCO observations also are in close agreement. 

We have identified a shock front at the southern edge 
of the CME.  A detailed analysis shows that the appearance of the 
CME-driven-shock in a synthetic coronagrpah image closely matches 
the same structure found the LASCO C3 image at 12:18 UT on October 28.
Both the rigor of this quantitative comparison and the fidelity in
reproducing coronagraph images of CMEs is unprecedented.
This is all the more remarkable in that there was no parameter fitting 
to produce the CME images, as only the CME speed at $20\,\RS$  was used as
a guide in specifying the flux rope.

The simulation allows the opportunity to view the October 28 CME as it 
would appear at large elongation in the HI2 coronagraph \citep[]{Lugaz:2005}.
Here, we found complex time evolution of the CME in the HI2 field of 
view as dense structures were temporarily highlighted as they came in
close proximity to the Thomson sphere.  These results vividly illustrate
the effects of CME propagation viewed at large elongation pointed out by
\citet{Vourlidas:2006}.  

An aspect of the simulation that can be improved upon is the speed of 
the CME, which is a factor of two too fast inside of $5\,\RS$, and does
not recover the measured velocity until $20\,\RS$. 
We conclude that most of the errors in CME speed originate from
the {\it ad hoc} initiation with a flux rope in a state of force
imbalance.  However, the expansion of the flux rope, in the realistic
3D coronal model, is remarkable in its ability to reproduce the
appearance of the CME in Thomson scattered white light. 
This result suggests that much of the density structure of the
CME is determined by the plasma distribution in the corona
surrounding the active region. 

Future models of CME initiation will be improved by the use 
vector magnetogram data rather than just fitting flux ropes to
the observed line-of-sight component of the field.  For this event,
vector magnetograms clearly show that the magnetic field of AR 10486
is very highly sheared with the field running nearly parallel
to the neutral line from which the CMEs erupt \citep[]{Liu:2006}.
The evolution of magnetic shear is made manifest by observations
of photospheric proper motions, which show strong shear flows along
the magnetic neutral line prior to CMEs \citep[]{Yang:2004, Deng:2006}.
These observations strongly suggest that the magnetic shear is ultimately
driving these CMEs.  A physical explanation and simulation of such shear
flows was first provided respectively by \citet{Manchester:2000} and
\citet{Manchester:2001}, which shows
that the flows are driven by the Lorentz force that occurs as the
magnetic field emerges in a stratified atmosphere.  Furthermore,
these Lorentz-force-driven flows may persist until the magnetic field 
becomes so highly sheared that there is a loss of equilibrium producing
eruptions that can drive CMEs \citet{Manchester:2003, Manchester:2004c}.
The need to resolve the photospheric pressure scale height has limited
these simulations to Cartesian domains.  Currently, we are adressing
the challenge to include this self-consistent shearing mechanism 
in a global model of CME initiation that is guided by vector magnetic
field observations.

\acknowledgments
This work has been supported in part by NASA AISRP grant NNG04GP89G, 
NASA STEREO grant NAS5-3131,
NSF ITR project ATM-0325332, and by DoD MURI grant F49620-01-1-0359.
W. Manchester and M. Opher are thankful for support provided by
LWS grant NNX06AC36G.   G. To\'th has been partially supported by
the Hungarian Science Foundation (OTKA, grant No.~T047042).  
We acknowledge the use of data from SOHO C2 and C3 coronagraphs.  
We gratefully acknowledge the supercomputing resources provided by NASA's
Columbia system under award  SMD1-Dec04-0099.
(OTKA, grant No.~T047042).



\clearpage

\begin{thebibliography}{}

\bibitem[Amari et al.(2003)] {Amari:2003} Amari, T., Luciani, J.~F., Aly, J.~J., Mikic, Z., \& Linker, J. 2003, \apj , 585, 1073

\bibitem[Billings(1966)] {Billings:1966} Billings, D. W., 1966, A guide to the
solar corona, Academic Press, San Diego, Calif.

\bibitem[Cohen et al.(2007)] {Cohen:2007a} Cohen, O., Sokolov, I.~V., Roussev, I.~I., Lugaz, N., Manchester, W.~B., Gombosi, T.~I. \& Arge, C.~N. 2007, JASTP, {\it in press}

\bibitem[Cohen et al.(2007)] {Cohen:2007b} Cohen, O., Sokolov, I.~V., Roussev, I.~I., Arge, C.~N., Manchester, W.~B., Gombosi, T.~I., Frazin, R.~A., Park, H., Butala, M.~D., Kamalabadi, F., \& Velli, M. 2007, \apj, 654, L163

\bibitem[Deng et al.(2006)] {Deng:2006} Deng, N., Xu, Y., Yang, G., Cao, W., Liu, C., Rimmele, R., Wang, H., \& Denker, C. 2006, \apj, 644, 1278

\bibitem[Dryer et al.(2004)] {Dryer:2004} Dryer, M., Smith, Z., Fry, C.~D., Sun, W., Deehr, C.~S. \& Akasofu, S.-I. 2004, Space Weather, 2, S09001, doi:10.1029/2004SW000087

\bibitem[Gombosi et al.(2001)] {Gombosi:2001} Gombosi, T.~I., T\'oth, G., De Zeeuw, D.~L., Hansen, K., Kabin, K., \& Powell, K.~G. 2001, JCP, 177, 176

\bibitem[Groth et al.(2000)] {Groth:2000} Groth, C.~P.~T., De Zeeuw, D.~L., Gombosi, T.~I. \& Powell, K.~G. 2000, JGR, 105, 25053

\bibitem[Howard et al.(2007)] {Howard:2007} Howard, R.~A. et al. 2007, Space Sci. Rev. 105, 25053

\bibitem[Intriligator et al.(2005)] {Intriligator:2005} Intriligator, D.~S., Sun, W., Dryer, M., Fry, C.~D., Deehr, C., \& Intriligator, J. 2005, JGR, 110, A09S10, doi:10.1029/2004JA010939

\bibitem[Jacobs et al.(2007)] {Jacobs:2007} Jacobs, C., van der Holst, B., \& Poedts, S. 2007, A\&A, 470, 359, doi:10.1051/0004-6361:20077305

\bibitem[Jackson et al.(2006)] {Jackson:2006} Jackson, B.~V., Buffington, A., Hick, P.~P., Wang, X., \& Webb, D. 2006, JGR, 111, A04S91, doi10.1029/2004JA010942

\bibitem[Krall et al.(2006)] {Krall:2006} Krall, J., Yurchyshyn, V.~B., Slinker, S., Skoug, R.~M., \& Chen, J. 2006, \apj, 642, 541

\bibitem[Koutchmy \& Lamy(1985)] {Koutchmy:1985} Koutchmy, S., \& Lamy, P. L. 1985, Properties and Interactions of Interplanetary Dust, ed. R. H. Giese \& P. L. Lamy (Dordrecht: Reidel), 63 

\bibitem[Liu et al.(2005)] {Liu:2005} Liu, C., Deng, N., Liu, Y., Falconer, D., Goode, P., Denker, C., \& Wang, H. 2005, \apj 622, 722

\bibitem[Liu \& Hayashi(2006)] {Liu:2006} Liu, Y., \& Hayashi, K. 2006, \apj 640, 1135

\bibitem[Lugaz et al.(2005)] {Lugaz:2005} Lugaz, L., Manchester IV, W.~B., \& Gombosi, T. 2005, \apj 627, 1019

\bibitem[Lugaz et al.(2007)] {Lugaz:2007} Lugaz, L., Manchester IV, W.~B., Roussev, I.~I., T\'oth, G.,  \& Gombosi, T. 2007, \apj 659, 788

\bibitem[Luhmann et al.(2004)] {Luhmann:2004} Luhmann, J. G., Solomon, S. C., Linker, J. A., Lyon, J. G., Mikic, Z., Odstrcil, D., Wang, W., \& Wiltberger, M. 2004, JATP, 66, 1243 

\bibitem[Manchester \& Low(2000)]{Manchester:2000} Manchester, IV,~W. 
\& Low, B.~C. 2001, Phys. of Plasmas, 7, 1263

\bibitem[Manchester(2001)]{Manchester:2001} Manchester, IV,~W. 2001,
\apj, 547, 503

\bibitem[Manchester(2003)]{Manchester:2003} Manchester, IV,~W. 2003, \jgr, 108, 1162, doi:10.1029/2002JA009252

\bibitem[Manchester et al.(2004a)] {Manchester:2004a} Manchester, W.~B., IV,
Gombosi, T.~I., Roussev, I., DeZeeuw, D.~L., Sokolov, I.~V., Powell, K.~G., 
T\'oth, G. \& Opher, M. 2004, JGR, 109, A01102, doi:10.1029/2002JA009672

\bibitem[Manchester et al.(2004b)] {Manchester:2004b} Manchester, W.~B., IV,
Gombosi, T.~I., Ridley, A.~J., Roussev, I., DeZeeuw, D.~L., Sokolov, I.~V., 
Powell, K.~G., \& T\'oth, G. 2004, JGR, 109, A02107, doi:10.1029/2003JA010150

\bibitem[Manchester et al.(2004c)]{Manchester:2004c} Manchester, IV,~W.,
Gombosi, T., DeZeeuw, D., \& Fan, Y. 2004, \apj, 610, 588

\bibitem[Mancuso et al.(2002)] {Mancuso:2002} Mancuso, S., Raymond, J.~C., Kohl, J., Ko, Y.-K., Uzzo, M., \& Wu, R. 2002, A\&A, 383, 267

\bibitem[Odstrcil (2003)] {Odstrcil:2003} Odstrcil, D. 2003, Adv. Space Res., 32, 497

\bibitem[Odstrcil et al.(2005)] {Odstrcil:2005} Odstrcil, D., Pizzo, V.~J., \& Arge, C.~N. 2005, JGR, 110, A02106, doi:10.1029/2004JA010745

\bibitem[Powell et al.(1999)] {Powell:1999} Powell, K.~G., Roe, P.~L., Linde, T.~J., Gombosi, T.~I.,\& De Zeeuw, D.~L. 1999, JCP, 154, 284

\bibitem[Raymond et al.(2000)] {Raymond:2000} Raymond, J. C., Thompson, B.~J., St. Cyr, O.~C., Gopalswamy, N., Kahler, S., Kaiser, M., Lara, A., Ciaravella, A., Romoli, M., \& O\'Neal, R. 2000, GRL, 1439.

\bibitem[Riley et al.(2002)] {Riley:2002} Riley P., Linker, J.~A., Miki\'c, Z., Odstrcil, D., Pizzo, V.~J., \& Webb, D.~F. 2002, ApJ 578, 972

\bibitem[Roussev et al.(2003a)] {Roussev:2003a} Roussev I.~I., Forbes, T.~G., Gombosi, T.~I. Sokolov, I.~V., De Zeeuw, D.~L., \&  Birn, J., 2003, ApJ 588, L457

\bibitem[Roussev et al.(2003b)] {Roussev:2003b} Roussev I.~I., Gombosi, T.~I. Sokolov, I.~V., Velli, M., Manchester, W., De Zeeuw, D.~L., Liewer, P., \&  T\'oth, G., 2003, ApJ 595, L57

\bibitem[Saito et al.(1977)] {Saito:1977} Saito, K., Poland, A.~I., \& Munro, R. 1977, Solar Phys., 55, 121

\bibitem[Skoug et al.(2004)] {Skoug:2004} Skoug, R.~M., Gosling, J.~T., Steinberg, J.~T., McComas, D.~J., Smith, C.~W., Ness, N.~F., Hu, Q., \& Burlaga, L.~F. 2004, JGR, 109, A09102

\bibitem[Sime \& Hundhausen(1987)] {Sime:1987} Sime, D. G. \& Hundhausen, A.~J., 1987, JGR, 92, 1049.

\bibitem[Su et al.(2006)] {Su:2006} Su, Y.~N., Golub, L., Van Ballegooijen, A.~A., \& Gros, M. 2006, Sol. Phys., 236, 325

\bibitem[Titov \& D\'emoulin(1999)]{Titov:1999} Titov, V.~S., \& D\'emloulin, P. 1999, A\&A, 351, 701

\bibitem[Tokumaru et al.(2007)]{Tokumaru:2007} Tokumaru, M., Kojima, M., Fujiki, K., Yamashita, M., \& Jackson, B. 2007, JGR, 112, A05106, doi:10.1029/2006JA012043

\bibitem[T\'oth et al.(2005)]{Toth:2005} Toth, G., et al. 2005, JGR,  110, A12226, doi10.1029/2005JA011126

\bibitem[T\'oth et al.(2006)]{Toth:2006} Toth, G., De Zeeuw, D.~L., Gombosi, T.~I., \& Powell, K. 2006, JCP, 217, 722 

\bibitem[T\'oth et al.(2007)]{Toth:2007} Toth, G., De Zeeuw, D., Gombosi, T., Manchester IV, W., Ridely, A., Sokolov, I., Roussev, I. 2007, Space Weather, 5, S06003 

\bibitem[Usmanov \& Dryer(1995)] {Usmanov:1995} Usmanov, A. V., \& Dryer, M. 1995, Sol Phys., 159, 347  

\bibitem[Usmanov et al.(2000)] {Usmanov:2000} Usmanov, A. V., Goldstein, M.~L., Besser, B.~P., \& Fritzer, J.~M. 2000, JGR, 105, 12675

\bibitem[Vourlidas et al.(2000)]{Vourlidas:2000} Vourlidas, A., Subramanian, Dere, K.~P., \& Howard, R.~A. 2000, \apj , 534, 456

\bibitem[Vourlidas et al.(2003)]{Vourlidas:2003} Vourlidas, A., Wu, S.~T., Wang, A.~H., Subramanian, P., \& Howard, R.~A. 2003, \apj , 598, 1392

\bibitem[Vourlidas \& Howard(2006)]{Vourlidas:2006} Vourlidas, A., \& Howard, R.~A. 2006, \apj, 642, 1216

\bibitem[Wang \& Sheeley(1990)]{Wang:1990} Wang, Y.-M., \& Sheeley, N.~R. 1990, \apj, 355, 726

\bibitem[Wu et al.(1999)] {Wu:1999} Wu, S. T., Guo, W.~P., Michels, D.~J.,
\& Burlaga L.~F. 1999, JGR, 14789, 1999

\bibitem[Wu et al.(2005)]{Wu:2005} Wu, C.-C., et al. 2005, JGR, 110, A09S17, doi:10.1029/2005JA011011

\bibitem[Yang et al.(2004)] {Yang:2004} Yang, G., Xu, Y., Cao, W., Wang, H., Denker, C., \& Rimmele, T.~R., 2004, /apj, 617, L151

\bibitem[Zurbuchen(2004)] {Zurbuchen:2004} Zurbuchen, T.~H., Gloeckler, G., Ipavich, F., Raines, J., Smith, C.~W., \& Fisk, L.~A.  2004, GRL, 31, L11805, doi:10.1029/2004GL019461

\end{thebibliography}

\end{document}